\documentstyle[jkas39]{article}

\runningauthor{S. C. KIM}
\runningtitle{OPEN CLUSTERS: NGC 1641 AND NGC 2394}
\beginpage{1}
\endpage{9}

\def\simlt{\lower.5ex\hbox{$\; \buildrel < \over \sim \;$}}
\def\simgt{\lower.5ex\hbox{$\; \buildrel > \over \sim \;$}}
\def\arcdeg{\hbox{$^\circ$}}
\def\arcmin{\hbox{$^\prime$}}
\def\arcsec{\hbox{$^{\prime\prime}$}}
\slugcomment{Not to appear in Nonlearned J., 45. \today}

\begin{document}

\title{NEAR-INFRARED PHOTOMETRIC STUDY OF THE GALACTIC OPEN CLUSTERS NGC 1641 AND NGC 2394 BASED ON 2MASS DATA}

\author{
{\normalsize\textbf{\textsc{S}}}\scriptsize\textbf{\textsc{ANG}}
{\normalsize\textbf{\textsc{C}}}\scriptsize\textbf{\textsc{HUL}}
{\normalsize\textbf{\textsc{K}}}\scriptsize\textbf{\textsc{IM}}
}

\offprints{S. C. Kim}
\address{Korea Astronomy and Space Science Institute, 61-1, Hwaam-dong, Yuseong-gu, 
  Daejeon 305-348, Korea}
\address{\it E-mail: sckim@kasi.re.kr}

\vskip 3mm
\address{\normalsize{\it (Received Oct. 26, 2006; Accepted ???. ??, 2006)}}

\abstract{
We present $JHK_S$ near-infrared CCD photometric study
  for the Galactic open clusters NGC 1641 and NGC 2394.
These clusters have never been studied before, 
  and we provide, for the first time, the cluster parameters; 
  reddening, distance, metallicity and age.
NGC 1641 is an old open cluster with age $1.6 \pm 0.2$ Gyr,
  metallicity [Fe/H]$=0.0 \pm 0.2$ dex, 
  distance modulus ${\rm (m-M)}_0 = 10.4 \pm 0.3$ mag (d$=1.2 \pm 0.2$ kpc), and
  reddening $E(B-V) = 0.10 \pm 0.05$ mag.
The parameters for the other old open cluster NGC 2394 are estimated 
  to be age = $1.1 \pm 0.2$ Gyr,
  [Fe/H]$=0.0 \pm 0.2$ dex,
  ${\rm (m-M)}_0 = 9.1 \pm 0.4$ mag (d$=660 \pm 120$ pc), and
  $E(B-V) = 0.05 \pm 0.10$ mag.
The metallicities and distance values for these two old open clusters
  are consistent with 
  the relation between the metallicities and the Galactocentric distances
  of other old open clusters. 
We find the metallicity gradient of 53 old open clusters 
  including NGC 1641 and NGC 2394
  to be $\Delta$[Fe/H]/$\Delta R_{gc}= -0.067 \pm 0.009$ dex kpc$^{-1}$.
}


\maketitle
\section{INTRODUCTION}
The Milky Way Galaxy has more than 1800 known open star clusters 
  and most of these clusters reside in the Galactic disk,
  while Piskunov et al. (2006) estimate that the total number of
  open clusters (OCs) in the Galactic disk to be of order of $10^5$
  at present.
OCs, therefore, are very important tools to study
  the formation and evolution of the Galactic disk and
  the star clusters themselves, and 
  the stellar evolution.
The fundamental physical parameters of open clusters,
  such as distance, interstellar reddening, chemical composition,
  age and metallicity,
  are necessary for the study of the clusters and the Galactic disk.
The Galactic (radial and vertical) abundance gradient also
  can be studied by open clusters
  (Hou, Prantzos, \& Boissier 2000; Chen, Hou, \& Wang 2003;
  Kim \& Sung 2003; Kim et al. 2005).
 
Despite the large number of open clusters in the Galaxy,
  much of them were discovered quite recently.
While the Lyng\r{a} (1987) ``Catalog of Open Cluster Data'' (COCD) 
  lists about 1200 clusters, recent discoveries of new open clusters
  including Dias et al. (2002, http://www.astro.iag.usp.br/$\sim$wilton/,
  version 2.7, 2006 October 27, 1759 clusters), Bica, Dutra, \& Barbuy (2003a),
  Dutra et al. (2003), Bica et al. (2003b),
  Alessi, Moitinho, \& Dias (2003), Borissova et al. (2003), 
  Bica, Bonatto, \& Dutra (2004),
  Kharchenko et al. (2005b),
  Mercer et al. (2005) and Kronberger et al. (2006)
  make the number of Galactic open clusters well over 1800.
Very recently, Froebrich, Scholz, \& Raftery (2006) have identified 
  1021 new (out of 1788 total) star cluster candidates only at $|b| < 20\arcdeg$.

Most of the above OCs and OC candidates were discovered very recently
  and lack detailed photometric studies.
This situation is almost the same even for the clusters that have
  been known to human for decades.
Only for about 400 out of $\sim 1200$ COCD clusters (Lyng\r{a} 1987)
  we have accurate, but heterogeneous $UBV$ photometry, and
  photometric distances, reddening and age values (Piskunov et al. 2006).
Dias et al. (2002)'s version 2.7 catalog gives the following statistics:
  out of 1759 OCs, only for 148 (8.41\%) clusters we have abundance values,
  for 316 (17.96\%) clusters we have distance, age, proper motion and 
    radial velocity values, and
  for 862 (49.00\%) clusters we have distance, reddening and age values.
The fact that there are more clusters to be studied than the clusters
  with studies already performed might be due to the higher speed of
  discovery of new clusters than that of study of each known cluster.
Some overall properties of the Galactic OCs have been recently discussed by
  Piskunov et al. (2006), Kharchenko et al. (2005a), and von Hippel (2005).

Recently the Two Micron All Sky Survey (Skrutskie et al. 1997, 
  Skrutskie et al. 2006, 2MASS, available at 
  http://www.ipac.caltech.edu/2mass/releases/allsky/)     \\
  including the Point-Source Catalogue and Atlas, 
  has produced huge amounts of near-infrared data
  (25.4 Tbytes of raw imaging data) covering 99.998\% of the celestial sphere
  from the observations between 1997 June and 2001 February.
In this paper, we have used the 2MASS photometry data to study 
  the previously unstudied open clusters NGC 1641 and NGC 2394.
The preliminary results on NGC 1641 presented in a conference proceedings
  (Kim 2006) are superseded by this paper.

Section II describes the near-infrared data.
Section III and IV present the analyses for NGC 1641 and NGC 2394,
  respectively.
Section V discusses the results and 
  a summary and conclusions are given in Section VI.

\section{THE 2MASS DATA}
The 2MASS project (Skrutskie et al. 2006) have used 
  two dedicated 1.3 m diameter telescopes 
  located at Mount Hopkins, Arizona, and Cerro Tololo, Chile and 
  $256 \times 256$ NICMOS3 (HgCdTe) arrays manufactured 
  by Rockwell International Science Center (now Rockwell Scientific),
  which give field-of-view of $8.'5 \times 8.'5$ and pixel scale of 
  $2\arcsec$ pixel$^{-1}$.
The photometric system comprise $J$ (1.25 $\mu$m),
  $H$ (1.65 $\mu$m) and $K_S$ (2.16 $\mu$m) bands, where 
  the ``$K$-short'' ($K_S$) filter excludes wavelengths longward of 2.31 $\mu$m
  to reduce thermal background and airglow and includes wavelengths as short as
  2.00 $\mu$m to maximize bandwidth (see Figure 2 of Skrutskie et al. 2006 or
  Figure 7 of Bonatto, Bica, \& Girardi 2004
  for the transmission curves of the 2MASS filters; Carpenter 2001).

VizieR\footnote{http://vizier.u-strasbg.fr/viz-bin/VizieR?-source=2MASS}
  was used to extract $J, H,$ and $K_S$ 2MASS photometry data
  in circular areas centered on the two clusters.
The photometry data are extracted for $R \le 6\arcmin$ for NGC 1641
  and for $R \leq 4\arcmin$ for NGC 2394,
  where $R$ means radius.
Figure 1 displays the grey-scale images of NGC 1641 (upper panel) 
  and NGC 2394 (lower panel)
  as taken from the Digitized Sky Surveys (DSS),
  which shows that both objects are loose clusters.
The centers of NGC 1641 and NGC 2394 are estimated approximately
  to be at 
  R.A.(2000)$ = 04^h~ 35^m~ 32^s$,
  Decl.(2000) $ = -65\arcdeg~ 45\arcmin~ 00\arcsec$ and
  R.A.(2000)$=07^h~ 28^m~ 36^s$,
  Decl.(2000) $ =+07\arcdeg~ 05\arcmin~ 12\arcsec$,
  respectively.
The Galactic coordinates of NGC 1641 and NGC 2394 are
  $l = 277.\arcdeg20$, $b = -38.\arcdeg32$ and
  $l = 210.\arcdeg78$, $b = +11.\arcdeg47$, respectively.
The approximate radii of NGC 1641 and NGC 2394 are estimated 
  to be 6\arcmin~ and 4\arcmin, respectively.

\begin{figure}[p]
  \epsfxsize=7.5cm
  \epsfysize=7.5cm
  \centerline{\epsffile{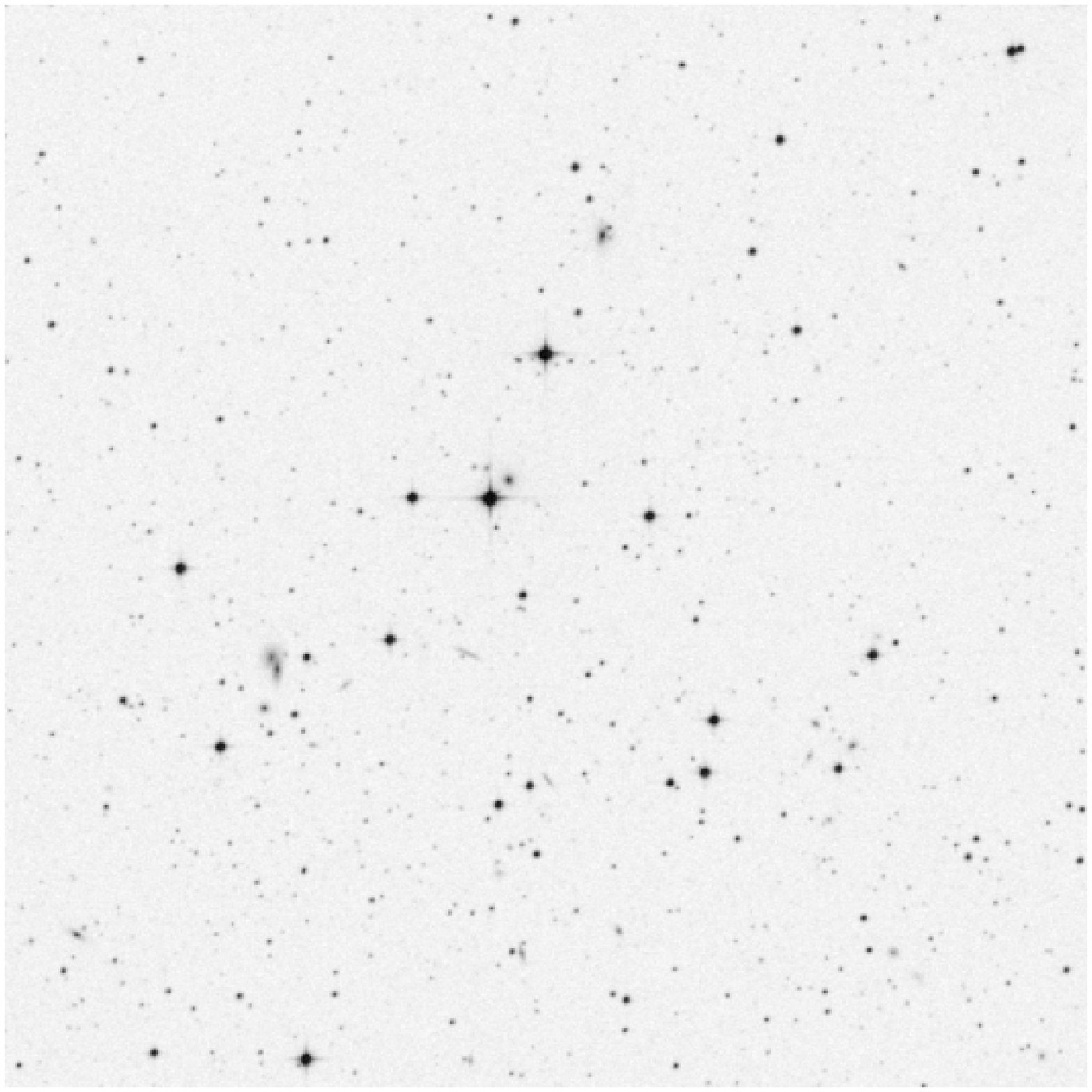}}
  \epsfxsize=7.5cm
  \epsfysize=7.5cm
  \centerline{\epsffile{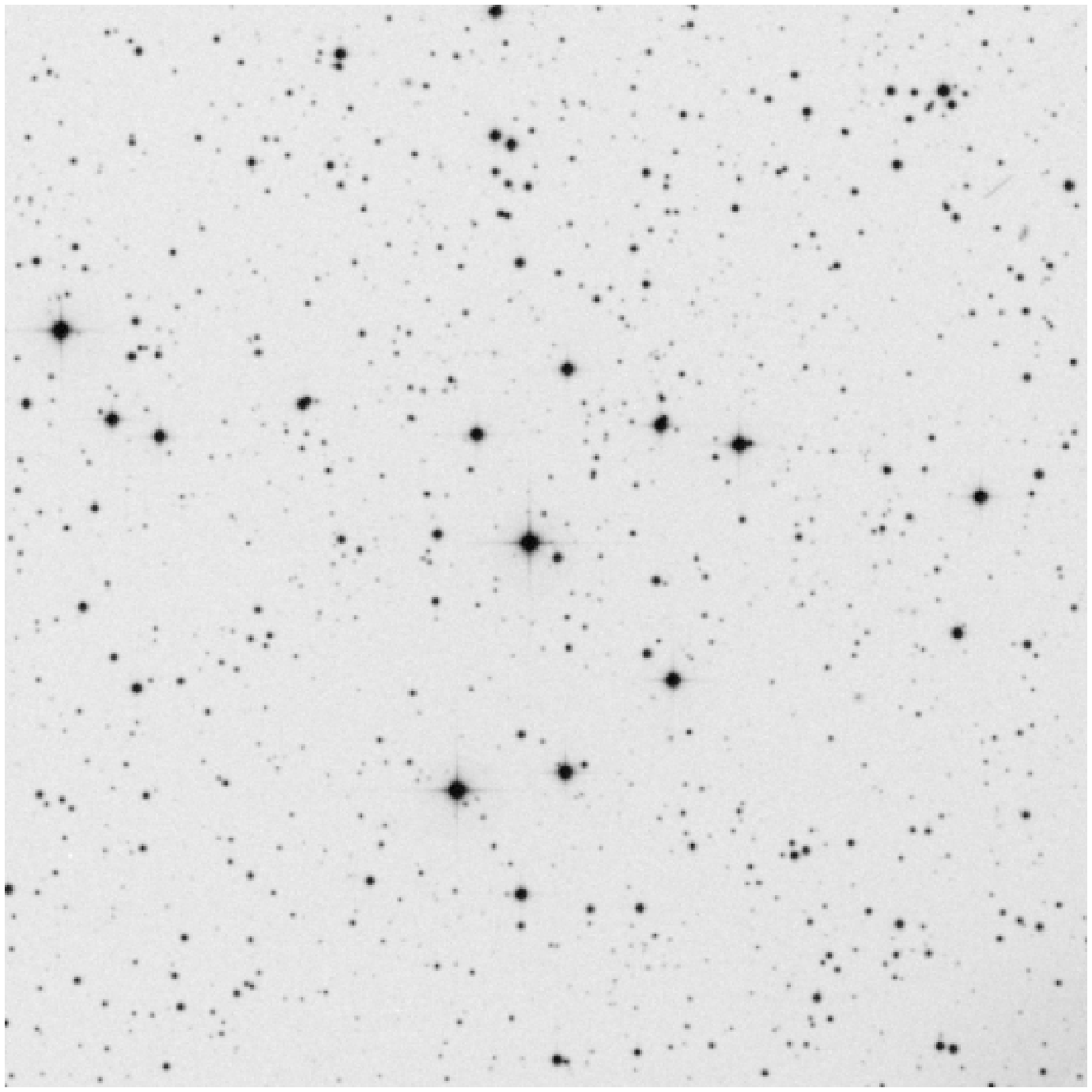}} 
{\small {\bf ~~~Fig. 1.}---~Grey-scale images of NGC 1641 (upper panel) 
  and NGC 2394 (lower panel)
  as taken from the Digitized Sky Surveys (DSS).
North is at the top and east is to the left.
The size of each field is $15\arcmin \times 15\arcmin$.
}
\end{figure}

\section{ANALYSIS FOR NGC 1641}
\subsection{The Color-Magnitude Diagrams}
Figure 2 shows the $J$ vs. $(J-H)$ (left panels) and 
  $K_S$ vs. $(J-K_S)$ (right panels) color-magnitude diagrams (CMDs)
  of stars in the region of NGC 1641 (upper panels), offset field 1 (middle panels)
  and offset field 2 (lower panels).
For each field, objects in the radius of 6\arcmin~ are used to draw 
  the CMDs.
Two offset fields are chosen at the same Galactic latitude, 
  but with one degree larger and one degree smaller Galactic longitude
  for offset field 1 and offset field 2, respectively, 
  than that of NGC 1641.

Although there are some possible field star contaminations at $J > 13$ mag
  or $K_S > 12.5$ mag, 
  it is clear that the region of NGC 1641 contains much larger
  number of stars at the brighter magnitude range,
  where the main sequence and possible 
  red giant branch or red giant clump (RGC)
  star(s) can be seen.

\begin{figure}
\centerline{\epsfxsize=8.0cm\epsfbox{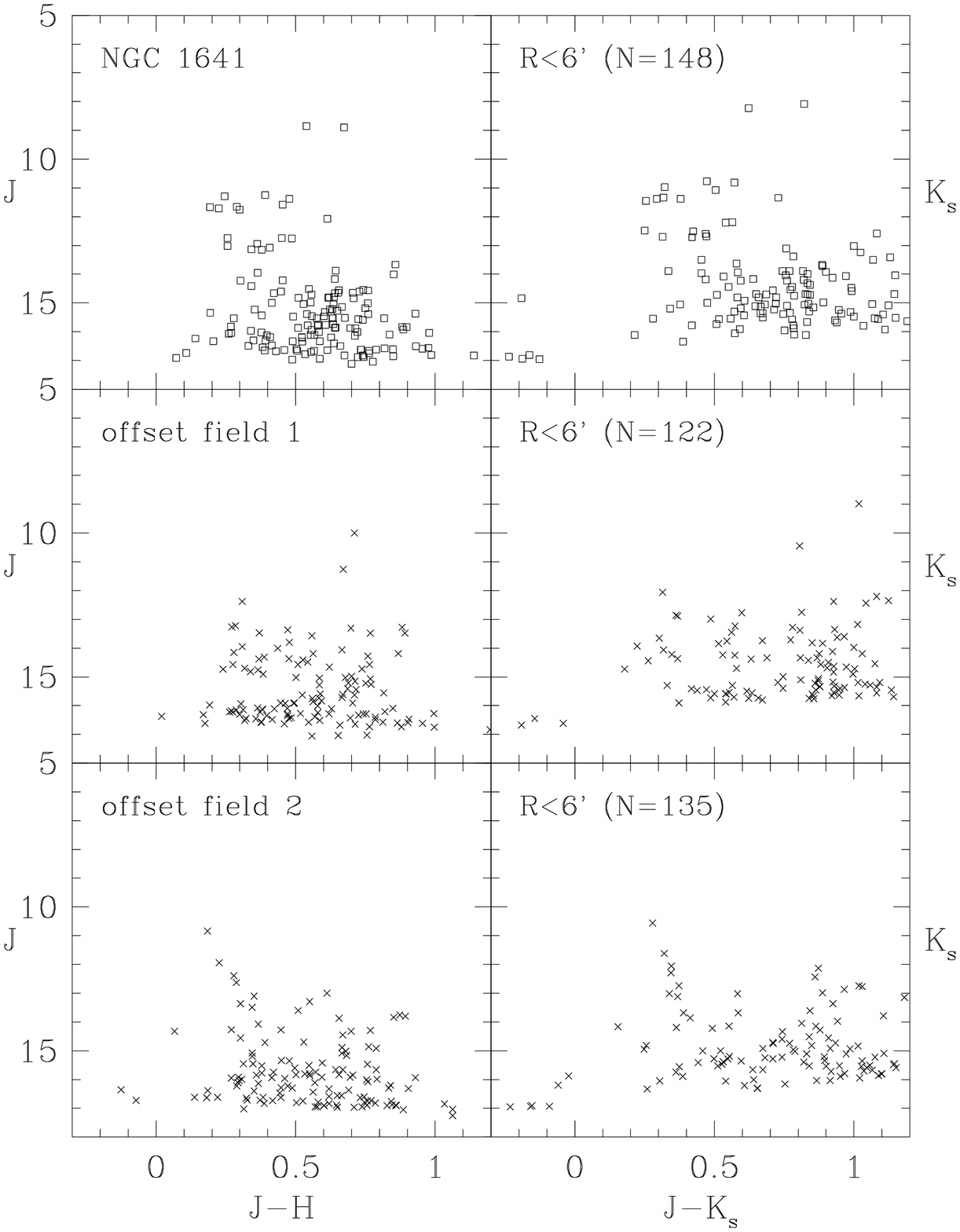}}
{\small {\bf ~~~Fig. 2.}---~$J$ vs. $(J-H)$ (left panels) and
  $K_S$ vs. $(J-K_S)$ (right panels) color-magnitude diagrams
  of the stars in NGC 1641 (upper panels), offset field 1 (middle panels)
  and offset field 2 (lower panels).
Extraction radius is 6\arcmin~ for each field.
Offset field 1 is the region with the same Galactic latitude and
  one degree larger Galactic longitude than that of NGC 1641, while
offset field 2 is the region with the same Galactic latitude and
  one degree smaller Galactic longitude than that of NGC 1641.
}
\end{figure}

\subsection{Padova Isochrone Fitting}
The near-infrared Padova isochrones of Girardi et al. (2000) have used
  the Johnson's $K$ band (Bessell \& Brett 1988, Cohen et al. 1992),
  while NGC 1641 and NGC 2394 have been observed and calibrated 
  with the 2MASS $K_S$ filter.
Therefore, we have fitted the new theoretical Padova isochrones 
  computed with the 2MASS
  $J, H$ and $K_S$ filters (Bonatto et al. 2004;
  Bica, Bonatto, \& Blumberg 2006)
  to derive the cluster parameters.
Figure 3 shows the best matched isochrones,
  which give the interstellar reddening of $E(B-V)=0.10 \pm 0.05$
  ($E(J-H)=0.309 E(B-V), E(J-K_S)=0.488 E(B-V)$), 
  the true distance modulus ${\rm (m-M)}_0 = (V-M_V) -3.1 \times E(B-V)
  = 10.4 \pm 0.3$ (d $= 1.2 \pm 0.2$ kpc), 
  metallicity [Fe/H] $= 0.0 \pm 0.2$ dex, and 
  age$= 1.6 \pm 0.2$ Gyr.
The color excess ratios we adopted (Bonatto et al. 2004) were derived 
  from absorption ratios in Schlegel, Finkbeiner, \& Davis (1998) and 
  the ratio $A_{K_S} / A_V =0.118$ from Dutra, Santiago, \& Bica (2002).

In Figure 4 we plotted a few more isochrones with different 
  ages (panels (a) and (b)) and metallicities (panels (c) and (d))
  for comparison.
Isochrones of twice the uncertainty smaller and larger age values
  are plotted as blue and green dotted lines, respectively, in panels (a) and (b), and
isochrones of one grid smaller and larger
  metallicity values are plotted as blue and green dotted lines, 
  respectively, in panels (c) and (d).
The uncertainties for the ages and metallicities of NGC 1641
  are estimated from this comparison.

\begin{figure}

\centerline{\epsfxsize=7.1cm\epsfbox{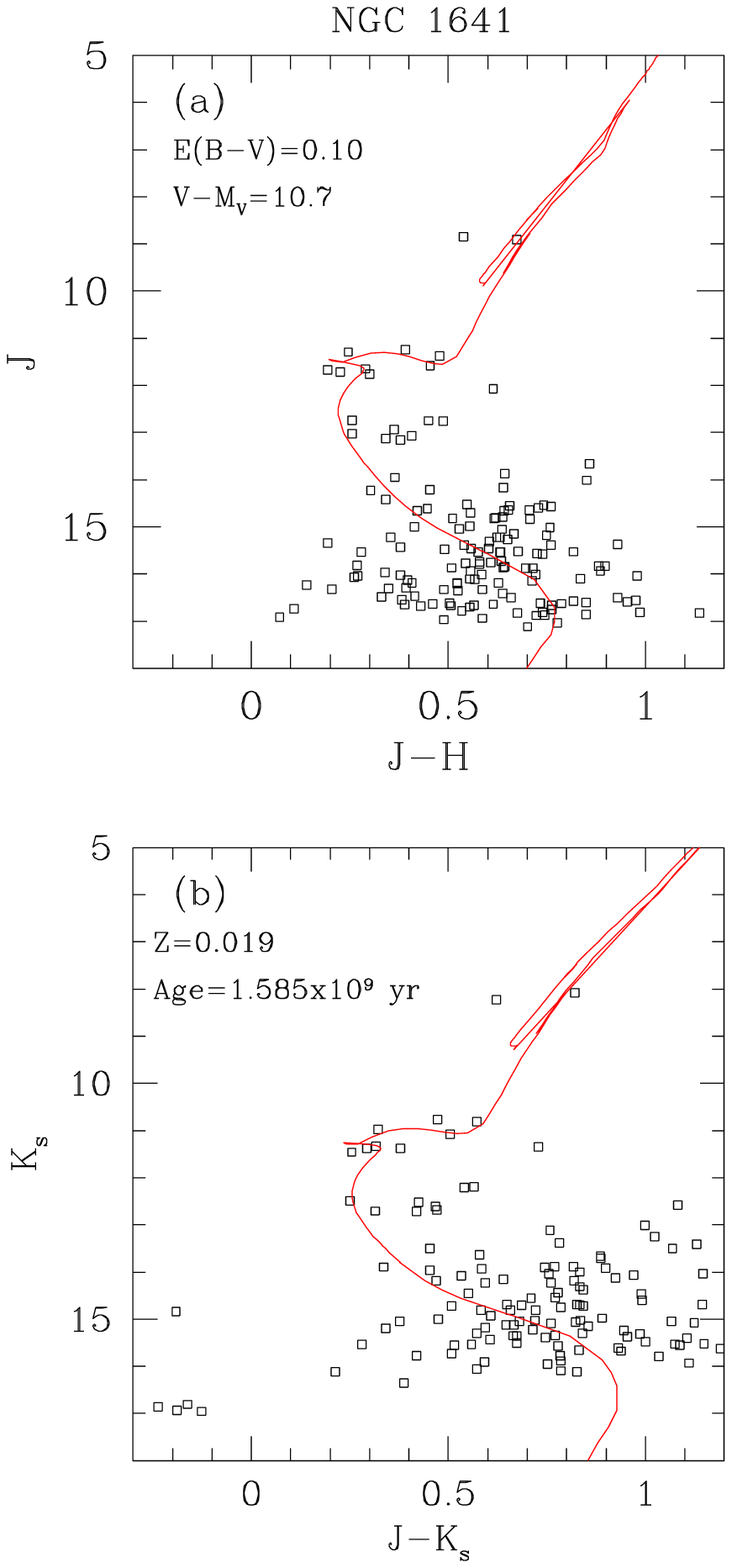}}
{\small {\bf ~~~Fig. 3.}---~Padova isochrone fittings for NGC 1641
  in (a) the $J$ vs. $(J-H)$ and (b) $K_S$ vs. $(J-K_S)$ color-magnitude diagrams.
The solid lines represent the Padova isochrones for 
  $E(B-V)=0.10$, $V-M_V = 10.7$,
  Z$=0.019$, and age = 1.6 Gyr.
}
\end{figure}

\begin{figure}

\centerline{\epsfxsize=9.1cm\epsfbox{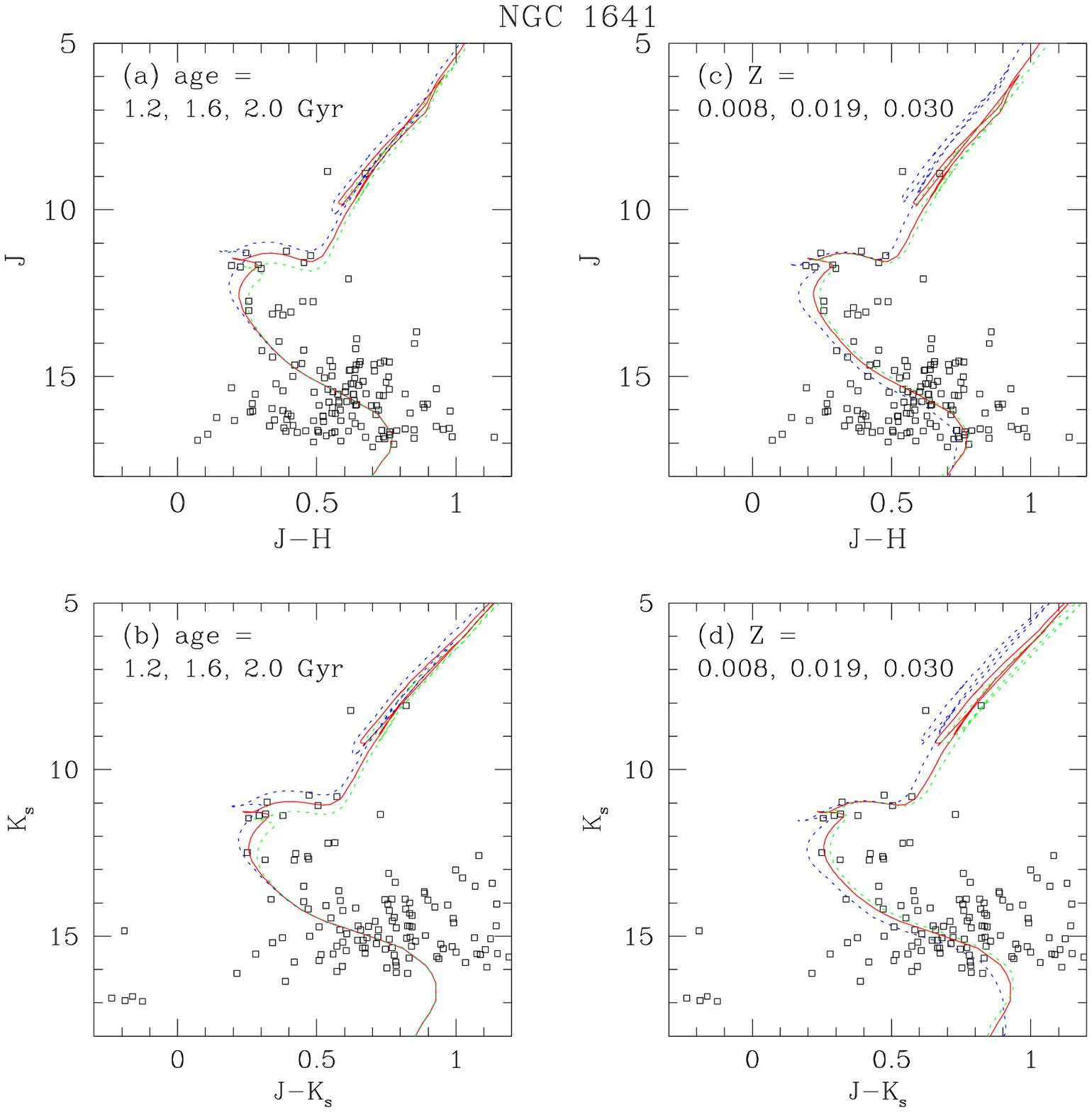}}
{\small {\bf ~~~Fig. 4.}---~Padova isochrone fittings for NGC 1641
  with different ages (panels (a) and (b)) and 
  different metallicities (panels (c) and (d))
  which are shown for comparison.
In the panels (a) and (b), isochrones of twice the uncertainty smaller and larger
  age values are plotted as blue and green dotted lines
  (left and right of the central red solid line), respectively.
In the panels (c) and (d), isochrones of one grid smaller and larger
  metallicity values are plotted as blue and green dotted lines
  (left and right of the central red solid line), respectively.
}
\end{figure}

\section{ANALYSIS FOR NGC 2394}
\subsection{The Color-Magnitude Diagrams}
Figure 5 shows the $J$ vs. $(J-H)$ (left panels) and
  $K_S$ vs. $(J-K_S)$ (right panels) CMDs
  of stars in the region of NGC 2394 (upper panels), offset field 1 (middle panels)
  and offset field 2 (lower panels).
For each field, objects in the radius of 4\arcmin~ are used to draw
  the CMDs.
As in the case of NGC 1641, two offset fields are chosen 
  at the same Galactic latitude with NGC 2394,
  but with one degree larger and one degree smaller Galactic longitude
  for offset field 1 and offset field 2, respectively,
  than that of NGC 2394.

Although there are some possible field star contaminations at $J$ or $K_S$ 
  magnitudes fainter than 13 mag
  and offset field 2 has some more brighter stars than offset field 1,
  it is clear that the region of NGC 2394 contains larger
  number of stars at the brighter magnitude range.
The CMDs of the region of NGC 2394 show the main sequence and possible 
  red giant branch or RGC stars.

\begin{figure}
\centerline{\epsfxsize=8.0cm\epsfbox{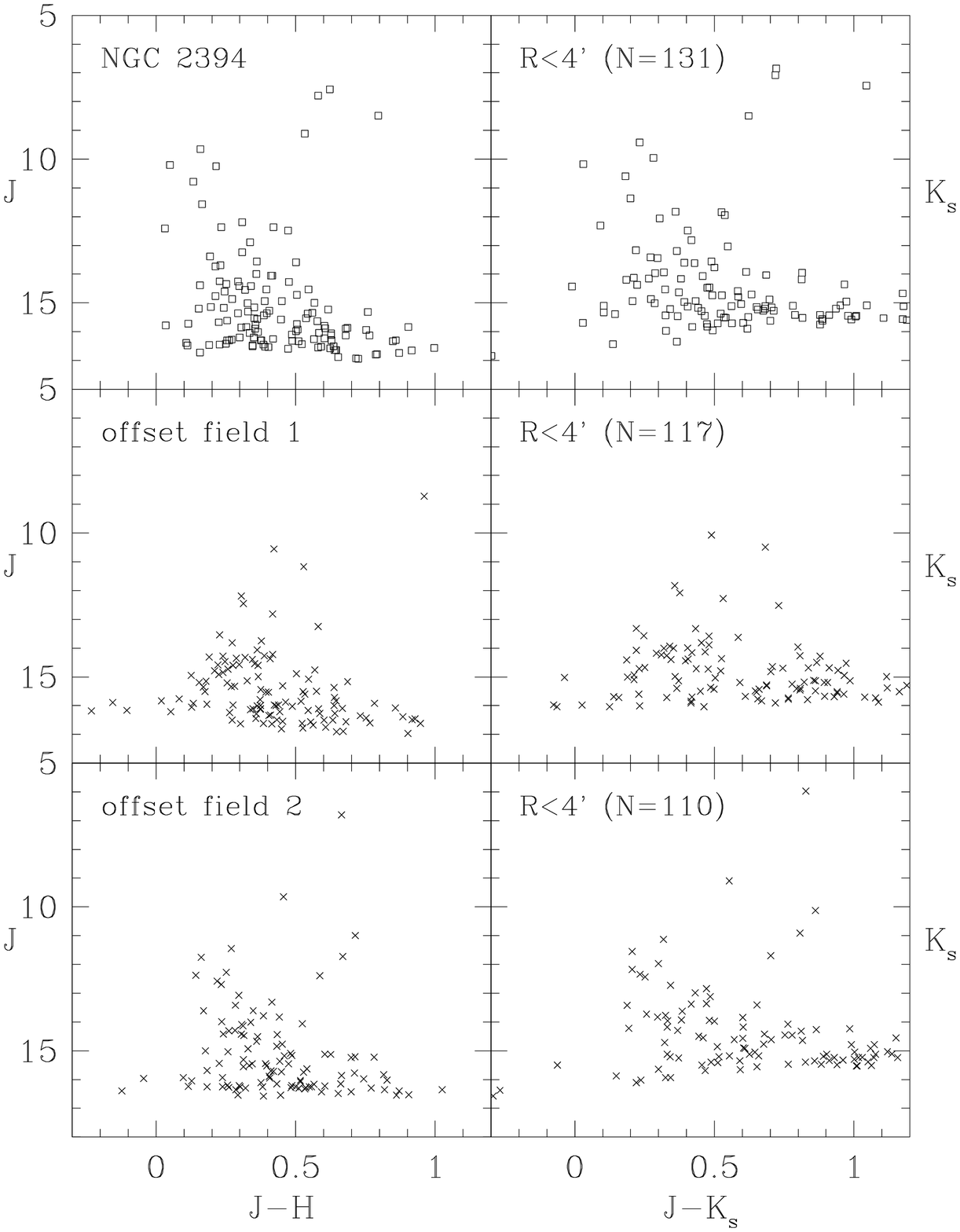}}
{\small {\bf ~~~Fig. 5.}---~$J$ vs. $(J-H)$ (left panels) and
  $K_S$ vs. $(J-K_S)$ (right panels) color-magnitude diagrams
  of the stars in NGC 2394 (upper panels), offset field 1 (middle panels)
  and offset field 2 (lower panels).
Extraction radius is 4\arcmin~ for each field.
Offset field 1 is the region with the same Galactic latitude and
  one degree larger Galactic longitude than that of NGC 2394, while
offset field 2 is the region with the same Galactic latitude and
  one degree smaller Galactic longitude than that of NGC 2394.
}
\end{figure}

\subsection{Padova Isochrone Fitting}
We have fitted the new theoretical Padova isochrones
  computed with the 2MASS $J, H$ and $K_S$ filters 
  to derive the parameters of NGC 2394 as in the case of NGC 1641.
Figure 6 shows the best matched isochrones,
  which give the interstellar reddening of $E(B-V)=0.05 \pm 0.10$,
  the true distance modulus ${\rm (m-M)}_0 = (V-M_V) -3.1 \times E(B-V)
  = 9.1 \pm 0.4$ (d $= 660 \pm 120$ pc), 
  metallicity [Fe/H] $= 0.0 \pm 0.2$ dex, and
  age$= 1.1 \pm 0.2$ Gyr.

The derived rather large age for an OC, and the existence of the red giant branch
  and sub giant branch stars
  imply that the star at $J=9.119$, $J-H=0.533$,
  $K_S=8.496$, and $J-K_S=0.623$ could be one of the RGC stars of NGC 2394.
Since the RGC, which are the stars of helium-burning stage, is 
  a typical feature of intermediate-aged and old OCs
  (Kronberger et al. 2006),
  a future deeper photometry of the cluster will be helpful
  for the confirmation of this.
Furthermore, RGC stars can be used as a standard candle 
  for the determination of a cluster's distance and reddening
  (Janes \& Phelps 1994; Phelps, Janes, \& Montgomery 1994).

In Figure 7 we plotted a few more isochrones with different
  ages (panels (a) and (b)) and metallicities (panels (c) and (d))
  for comparison.
Isochrones of twice the uncertainty smaller and larger age values
  are plotted as blue and green dotted lines, respectively, in panels (a) and (b), and
isochrones of one grid smaller and larger
  metallicity values are plotted as blue and green dotted lines,
  respectively, in panels (c) and (d).
The uncertainties for the ages and metallicities of NGC 2394 
  are estimated from this comparison.

\begin{figure}
\centerline{\epsfxsize=7.1cm\epsfbox{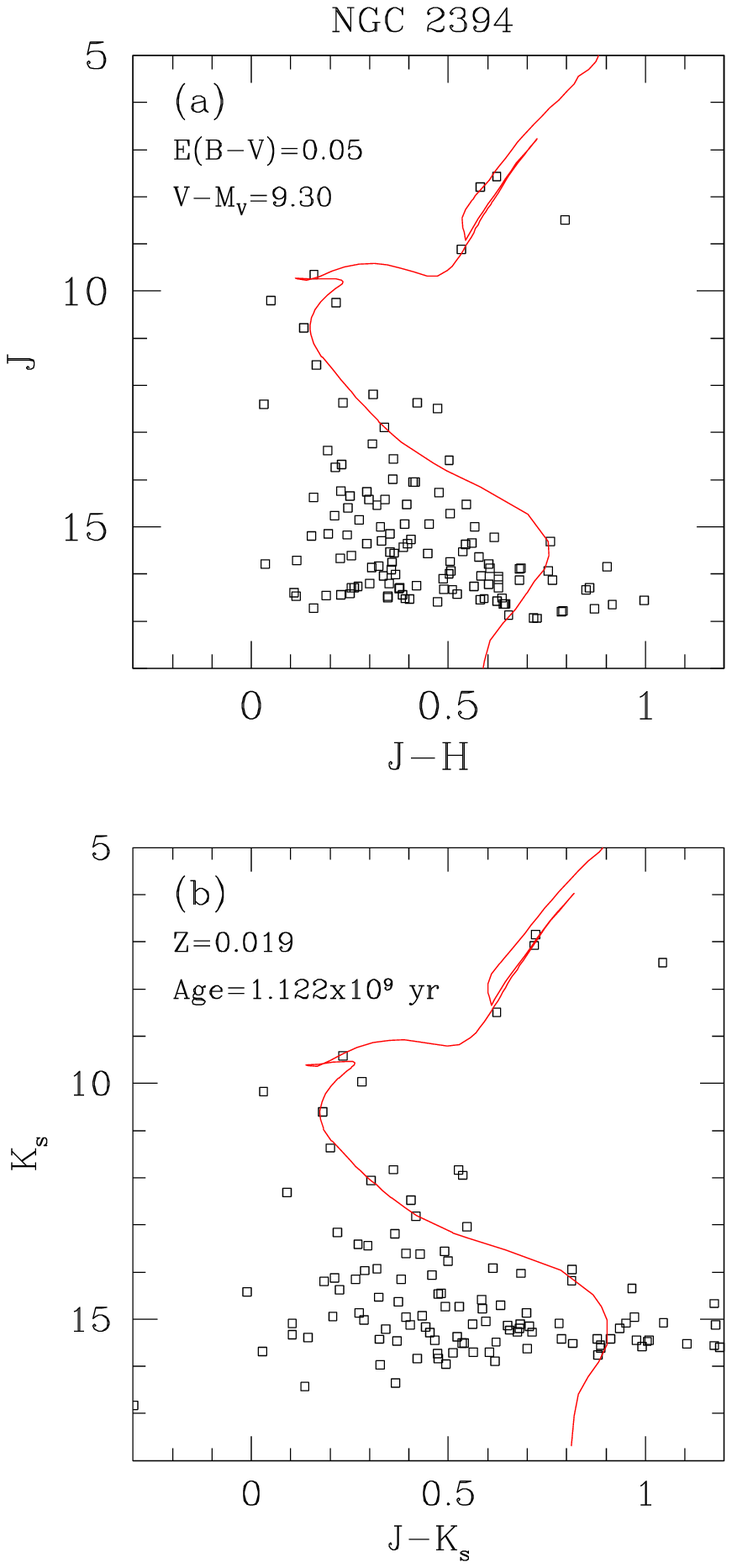}}
{\small {\bf ~~~Fig. 6.}---~Padova isochrone fittings for NGC 2394
  in (a) the $J$ vs. $(J-H)$ and (b) $K_S$ vs. $(J-K_S)$ color-magnitude diagrams.
The solid lines represent the Padova isochrones for
  $E(B-V)=0.05$, $V-M_V = 9.30$,
  Z$=0.019$, and age = 1.1 Gyr.
}
\end{figure}

\begin{figure}

\centerline{\epsfxsize=9.1cm\epsfbox{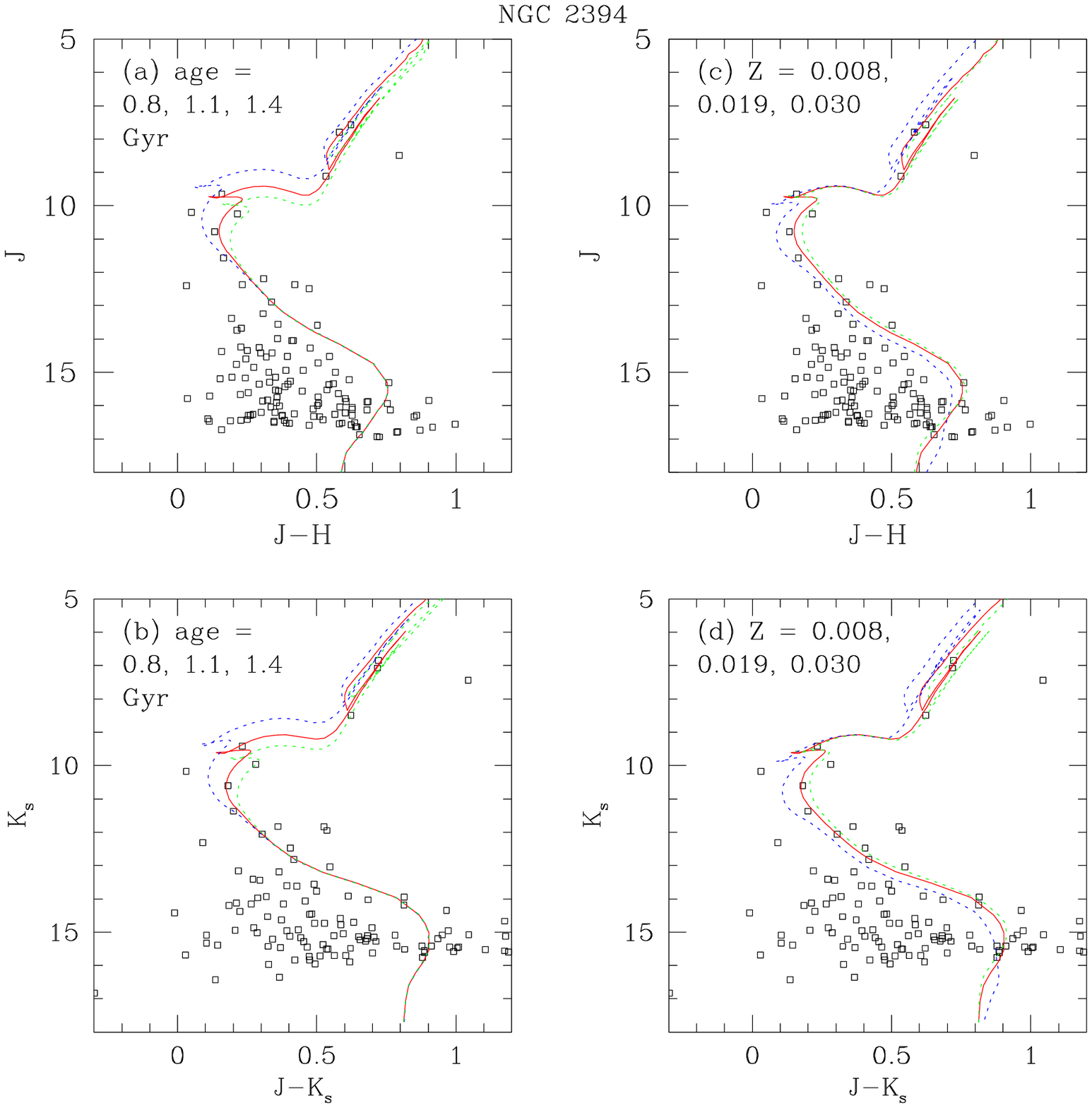}}
{\small {\bf ~~~Fig. 7.}---~Padova isochrone fittings for NGC 2394
  with different ages (panels (a) and (b)) and
  different metallicities (panels (c) and (d))
  which are shown for comparison.
In the panels (a) and (b), isochrones of twice the uncertainty smaller and larger
  age values are plotted as blue and green dotted lines
  (left and right of the central red solid line), respectively.
In the panels (c) and (d), isochrones of one grid smaller and larger
  metallicity values are plotted as blue and green dotted lines
  (left and right of the central red solid line), respectively.
}
\end{figure}

\section{DISCUSSION}
\subsection{Color-Color Diagrams}
Figure 8 shows the $(J-H) \times (H-K_S)$ color-color diagrams of 
  the stars with $K_S$ magnitude errors less than 0.1 
  in (a) NGC 1641 and (b) NGC 2394. 
The main sequence ranges of the Padova isochrones are overplotted
  as solid lines ($0.15 - 1.75 M_\odot$ for NGC 1641 and
  $0.15 - 1.55 M_\odot$ for NGC 2394) and
  the reddening vector $E(J-H) = 1.72 \times E(H-K_S)$ 
  for $A_V = 1$ are denoted as arrows.

Most of the stars in NGC 1641 and NGC 2394 are distributed along 
  the main sequences of the Padova isochrones and this suggests
  that the reddening values in the fields of these clusters
  are not so high.
The fact that stars in NGC 2394 are more tightly gathered around
  the Padova isochrones than those in NGC 1641 is consistent with
  the somewhat lower reddening value for NGC 2394 derived in Section IV
  than that for NGC 1641.

\begin{figure}
\centerline{\epsfxsize=7.1cm\epsfbox{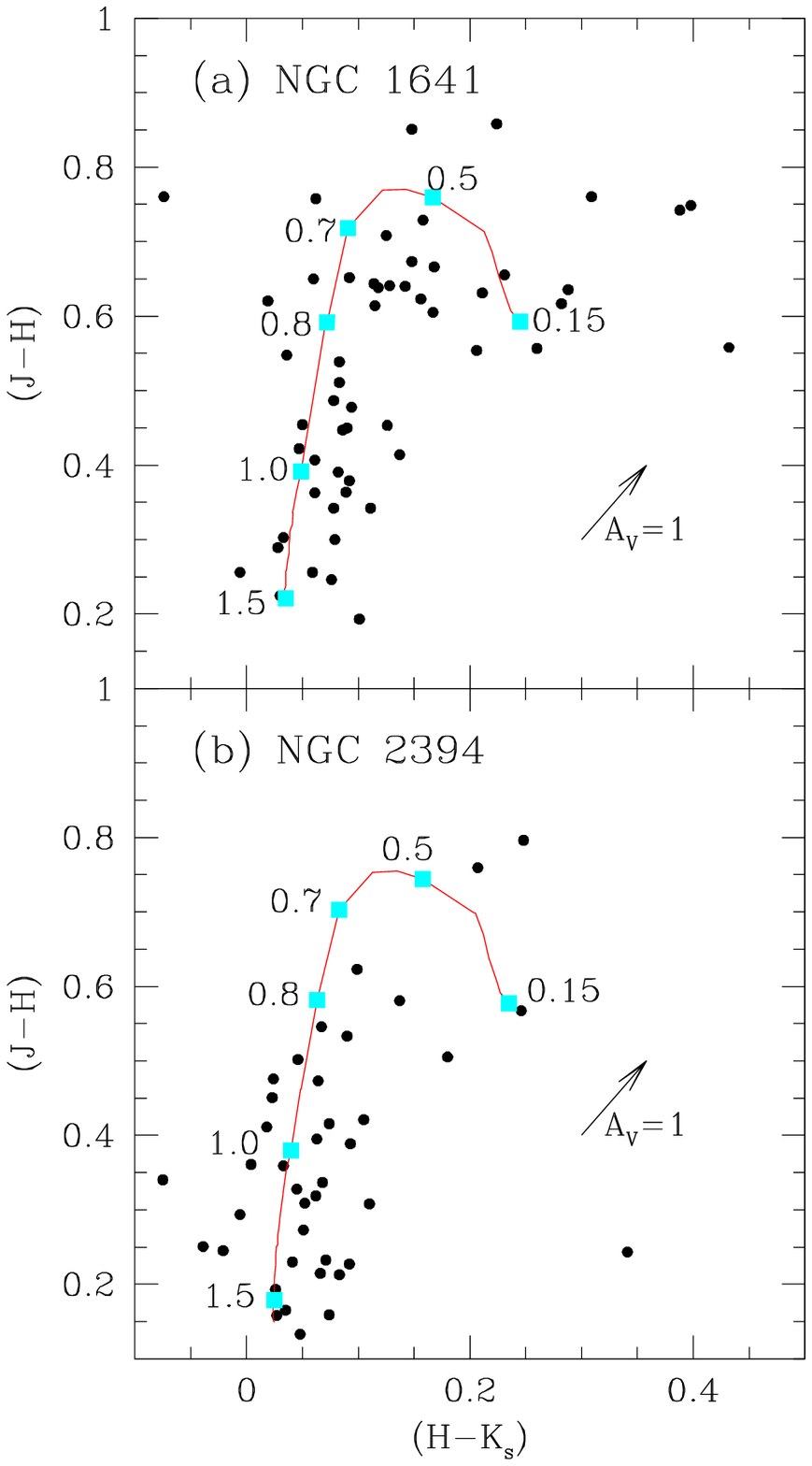}}
{\small {\bf ~~~Fig. 8.}---~Color-color diagrams of the stars in 
  (a) NGC 1641 and
  (b) NGC 2394 for $K_S$ magnitude errors less than 0.1.
Solid lines are main-sequence ranges of the Padova isochrones of ages 
  1.6 Gyr for NGC 1641 and 1.1 Gyr for NGC 2394.
Arrows are reddening vectors $E(J-H) = 1.72 \times E(H-K_S)$ for
  $A_V = 1$.
Representative masses in $M_\odot$ are indicated along the isochrones.
}
\end{figure}

\subsection{[Fe/H]--Galactocentric Radius Relation}
Using the Galactic coordinates and distances of NGC 1641 and NGC 2394
  from the Sun derived above, 
  we have calculated the Galactocentric distances of
  these clusters to be $8.4 \pm 0.01$ kpc and $9.1 \pm 0.1$ kpc,
  respectively (adopting the Galactocentric distance of the Sun
  as 8.5 kpc).
If we adopt the Galactocentric distance of the Sun of 8.0 kpc,
  these distances become $7.9 \pm 0.01$ kpc and $8.6 \pm 0.1$ kpc
  for NGC 1641 and NGC 2394, respectively.

Kim et al. (2005) have compiled the recent results of the various
  estimates of the slopes of the Galactocentric radial [Fe/H] gradient
  of old open clusters
  in their Table 2, whose mean value is $\Delta$[Fe/H]/$\Delta R_{gc}
  = -0.066 \pm 0.019$ dex kpc$^{-1}$.
They obtained a new value of the gradient $-0.064 \pm 0.009$ dex kpc$^{-1}$
  for 51 OCs including the cluster Czernik 24.

From including the current two old open clusters NGC 1641 and
  NGC 2394 and using the parameters obtained in this study
  in the plot of Figure 9, 
  we redetermined the gradient value to be $-0.067 \pm 0.009$ dex kpc$^{-1}$
  denoted as the solid line.
Again this value is in very good agreement with the mean values 
  obtained in Kim et al. (2005), and it can be said that
  the metallicities and the Galactocentric distances of 
  NGC 1641 and NGC 2394 obtained in this study are consistent
  with the general trend of old open clusters.

\begin{figure*}
\centerline{\epsfxsize=9.1cm\epsfbox{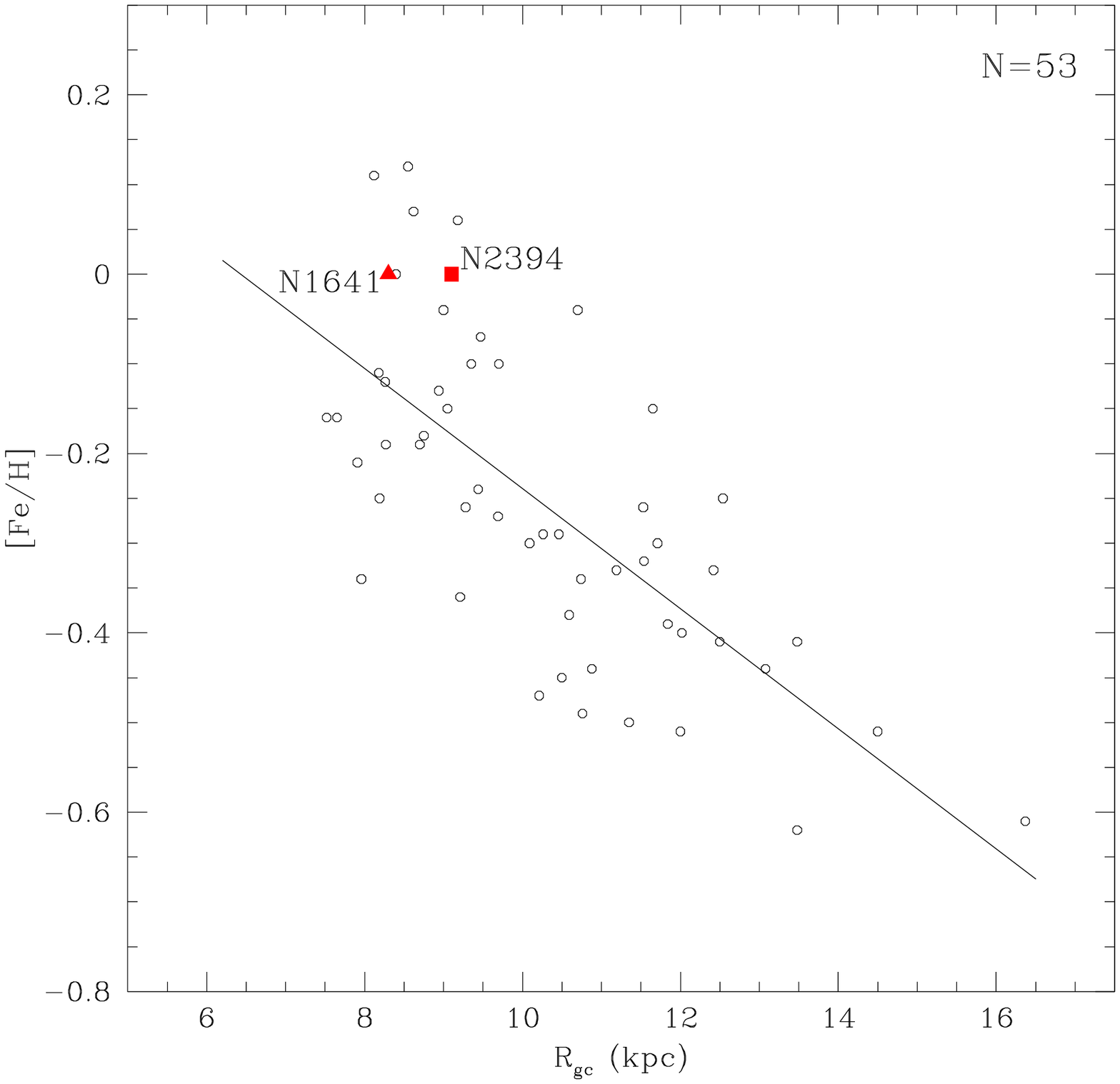}}
{\small {\bf ~~~Fig. 9.}---~Radial abundance gradient for 53 old open clusters
  including NGC 1641 (filled triangle) and NGC 2394 (filled square).
The solid line is a least-squares fit to the data that yields an
  abundance gradient of $\Delta$[Fe/H]/$\Delta R_{gc}=
  -0.067 \pm 0.009$ dex kpc$^{-1}$.
}
\end{figure*}

\section{SUMMARY AND CONCLUSIONS}
We have used the $JHK_S$ 2MASS near-infrared photometric data of 
  the old open clusters NGC 1641 and NGC 2394 
  to study these clusters photometrically and 
  determined the fundamental parameters of 
  ages, metallicities, distances, and interstellar reddening values. 
This is the first photometric study for these two clusters.
The resulting fundamental parameters for the two clusters are
  summarized in Table 1.
It is very anticipated to perform further optical photometric and/or
  deeper near-infrared photometric study for these clusters in the future.

\begin{table*}[t]
\begin{center}
{\bf Table 1.}~~Basic Information of NGC 1641 and NGC 2394\\
\vskip 3mm
{\small
\setlength{\tabcolsep}{1.2mm}
\begin{tabular}{lccl} \hline\hline
Parameter & NGC 1641 & NGC 2394 & Reference \\
\hline
$\alpha_{2000}$ & 04$^h$ 35$^m$ 32$^s$               & 07$^h$ 28$^m$ 36$^s$ & This study \\
$\delta_{2000}$ & $-65$\arcdeg~ 45\arcmin~ 00\arcsec & +07\arcdeg~ 05\arcmin~ 12\arcsec & This study \\
$l$ & 277.\arcdeg20                                  & 210.\arcdeg78 & This study \\
$b$ & $-38.\arcdeg32$                                & $+11.\arcdeg47$  & This study \\
Reddening, $E(B-V)$               & $0.10 \pm 0.05$ mag  & $0.05 \pm 0.10$ mag & This study \\
Distance modulus, $V_0 - M_V$     & $10.4 \pm 0.3$ mag   & $9.1 \pm 0.4$ mag  & This study \\
Distance, d                       & $1.2 \pm 0.2$ kpc    & $660 \pm 120$ pc   & This study \\
Galactocentric distance, $R_{gc}$ & $8.4 \pm 0.01$ kpc    & $9.1 \pm 0.1$ kpc & This study \\
Metallicity, Z                    & $0.019$              & $0.019$            & This study \\
Age, $t$                          & $1.6 \pm 0.2$ Gyr (${\rm log} ~t=9.20$)
                                  & $1.1 \pm 0.2$ Gyr (${\rm log} ~t=9.05$) & This study \\
\hline
\end{tabular}
} 
\end{center}
\end{table*}

\vspace{4mm}
We are grateful to the anonymous referee for numerous comments and
  suggestions that improved the quality of this manuscript.
This publication makes use of data products from the Two Micron All Sky Survey,
  which is a joint project of the University of Massachusetts and
  the Infrared Processing and Analysis Center/California Institute of Technology,
  funded by the National Aeronautics and Space Administration.
This research has made use of the SIMBAD database,
  operated at CDS, Strasbourg, France.


\end{document}